# Versatile optical frequency combs based on multi-seeded femtosecond optical parametric generation

Mikhail Roiz[1,*] and Markku Vainio[1,2,3]

[1]Department of Chemistry, University of Helsinki, FI-00560, Helsinki, Finland
[2]Photonics Laboratory, Physics Unit, Tampere University, Tampere, FI-33101, Finland
[3]e-mail: markku.vainio@helsinki.fi
*Corresponding author: mikhail.roiz@helsinki.fi

**Abstract:** This study proposes and demonstrates a versatile method for near and mid-infrared optical frequency comb generation using multi-seeded femtosecond optical parametric generation. The method allows one to divide the repetition rate by an arbitrarily large integer factor, freely tune the offset frequency, and adjust the common phase offset of the comb modes. Since all possible degrees of freedom are adjustable, the proposed method manifests itself as versatile optical frequency synthesis.

## 1. Introduction

Optical Frequency Comb (OFC) technology is a key metrological tool that combines unique properties, such as precision, coherence, and broad optical bandwidth that are vital for various scientific and industrial applications [1, 2]. A special type of OFCs – the fully stabilized OFC – phase-coherently links thousands of optical frequencies to radio frequency standards, which turns the OFC into an accurate optical frequency synthesizer or optical ruler [3-9].

The foundation for optical frequency comb generation has been laid by mode-locked lasers (MLLs), for which the electric field can be written

$$E(t) = \sum_{n=N_i}^{N_f} A_n\ e^{i2\pi n f_r t} e^{i2\pi f_{\text{CEO}} t} e^{i\varphi}. \tag{1}$$

Here, $A_n$ is the electric field amplitude and $n$ is an integer mode number that takes values from $N_i$ to $N_f$. The instantaneous phase of the field depends on two radiofrequencies, the carrier-envelope offset (CEO) frequency $f_{\text{CEO}}$ and the repetition rate $f_r$. In addition, we have included a frequency-independent phase offset $\varphi$ that we refer to as *common phase*. We deliberately separate the exponents to emphasize these three degrees of freedom in OFCs, which are all discussed in this paper. From (1) we also see that the optical frequencies of the frequency comb teeth can be expressed as [1, 2]:

$$\nu_n = f_{\text{CEO}} + n f_r. \tag{2}$$

An important prerequisite for precise optical frequency synthesis is CEO stabilization. With MLL frequency combs, this can be achieved using a technique called self-referencing or *f-2f* interferometry [10-14]. Self-referencing and other CEO stabilization techniques have also been demonstrated with electro-optic frequency combs [15], quantum-cascade lasers [16] and soliton microcombs [17-20], opening up access to repetition rates difficult to attain in MLLs.

Active and passive CEO stabilization methods alternative to *f-2f* interferometry have been established using various parametric down-conversion techniques, which are commonly used for mid-infrared (MIR) comb generation [6]. One such technique - Difference Frequency Generation (DFG) - leads to passive offset-free stabilization when driven by two frequency combs originating from the same laser [21-25]. In addition, it has been demonstrated that instead of passive CEO stabilization in DFG one can gain access to active control over the CEO through amplitude to phase modulation transfer in white-light generation that is used to seed the DFG process [26]. Due to the inherent feedback mechanism via optical cavity, Optical Parametric Oscillators (OPOs) also support fully stabilized OFC generation [27-31]. Furthermore, we have recently demonstrated that continuous-wave (CW)-seeded Optical Parametric Generation (OPG) allows for active stabilization and dynamic control of the MIR (idler) frequency comb CEO without stabilization of the pump CEO [32, 33]. Hence, the parametric down-conversion processes not only allow one to generate OFCs in new spectral regions but also to possess advantages in precise optical frequency synthesis.

Besides CEO stabilization and tuning, there is another important degree of freedom in OFCs that must be addressed – repetition rate or mode spacing $f_r$. This parameter is usually set by the cavity length in MLLs and soliton microcombs, and only the electro-optic frequency combs have the ability to set any arbitrary repetition rate, albeit at the cost of increased phase noise [34]. Repetition rate stabilization is relatively straightforward for MLL OFCs, provided that the optical cavity is equipped with piezoelectric actuators. Nowadays, even soliton microcombs have such piezoelectric actuators for active cavity stabilization and tuning [17]. However, the repetition rate tuning range is limited by the maximum travel of the piezo elements, or other elements used to control the cavity length. Therefore, much of research has been devoted to repetition rate multiplication and division. Well-known multiplication techniques are based on either filtering in optical cavities [35, 36] or temporal Talbot effect [37, 38]. Repetition rate division by an integer fraction has been demonstrated for MLLs using, for example, Pockels cells or acousto-optic modulators [39, 40]. This method is called pulse picking because pulses are periodically filtered out from the initial pulse train using a modulator. Since soliton microcombs usually have extremely high repetition rates, there have been attempts to divide the repetition rate using coupled microresonators [41, 42]. This method is limited to small division factors, however, since the size or number of microresonators increases when large division factors (>10) are desired. Overall, the repetition rate division is an important problem in OFC technology, because repetition rates well below 1 GHz are needed, for example, in high-resolution molecular spectroscopy. This is particularly true in the mid-infrared region, where the availability of high-speed photodetectors is limited.

In this article, we introduce and experimentally demonstrate a simple method for optical frequency comb generation with the capability of precisely adjusting all degrees of freedom shown in Eq. (1): $f_r$, $f_{CEO}$ and $\varphi$. The method is based on femtosecond OPG, which produces two combs - signal and idler - when pumped by a femtosecond laser. When the process is seeded with a CW laser locked to the pump, the CEO of the idler (or signal) comb can be accurately controlled, as we have previously shown [32, 33]. An advantage of this approach is that the comb offset frequency is inherently known; its measurement is not required. Here we place emphasis on the new multi-seeding approach, where instead of a single mode CW laser we use multiple lasers or modulation sidebands for seeding. It allows repetition rate division by, in principle, any arbitrarily large integer factor. Additionally, we show that the common phase $\varphi$ can be precisely controlled. We also discuss the CEO stabilization and tuning for completeness, although it was already studied in our previous work [32, 33]. Since all possible degrees of freedom are adjustable, the method presented in this paper can be considered as versatile optical frequency synthesis.

## 2. Principle

The basic idea of CW-seeded femtosecond OPG is illustrated in Fig. 1a. Femtosecond pulses from a mode-locked pump laser are focused into a second-order nonlinear crystal, such as lithium niobate. In non-degenerate OPG, the pump photons split into two lower energy photons called the signal and idler, and the pump repetition rate $f_r$ is also transferred to the signal and idler. Since the OPG process is based on quantum noise amplification, the absolute positions of the generated signal and idler combs vary randomly from pulse to pulse. In other words, their CEOs are random. However, if one introduces an additional source of photons in the signal region represented by a single-mode CW seed laser, the situation changes drastically. The seed serves as a pre-generated comb tooth for the signal comb, thus eliminating the pulse-to-pulse variations of the comb teeth positions (see Fig. 1a). This means that the CW seed laser defines the signal and idler comb CEOs [32, 33]. Such seeding also reduces the OPG threshold by up to 50%, depending on the input average power of the seed laser [32, 33, 43]. In order to avoid confusion in the terminology, we would like to point out that we call this method *CW-seeded OPG* instead of *Optical Parametric Amplification* (OPA) or DFG. Contrary to OPA and DFG, our OPG always has a threshold no matter if the seeding field is present or not [32, 33]. In addition, the bandwidth and overall shape of the signal spectrum is practically independent of the seeding. However, it is crucial to have a coherent seeding field to turn this light source into a *fully stabilized frequency comb*.

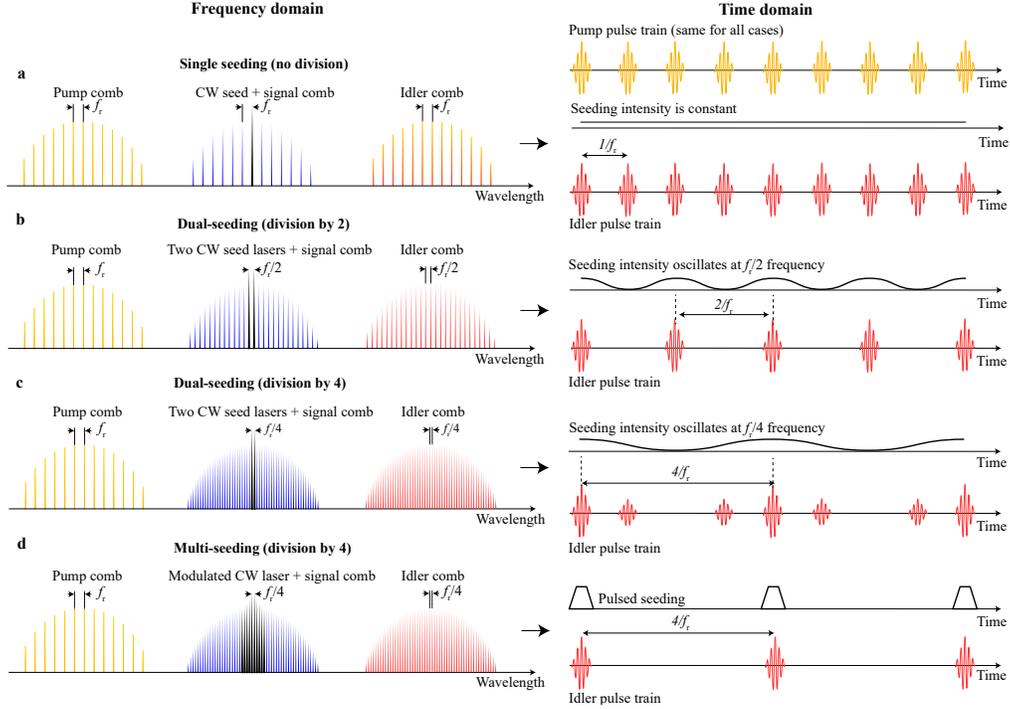

Fig. 1. Multi-seeded OPG concept. (a) OPG with single CW seeding. Note that idler pulses (same for signal) are generated for each pump pulse. (b) Dual-seeded OPG with mode spacing of $f_r/2$; the seeding field consists of two CW lasers optically beating at a frequency of $f_r/2$, hence at the moment of complete destructive interference there is not seeding field for each second pump pulse, so the idler pulses are not generated. (c) Dual-seeded OPG with mode spacing of $f_r/4$. Due to the additional frequency mixing new comb teeth are generated effectively making the idler and signal mode spacing equal to $f_r/4$. However, in time domain the pulse intensities are modulated. (d) Multi-seeded OPG with mode spacing of $f_r/4$. Note that since the seeding field is pulsed with repetition frequency of $f_r/4$, the OPG can start up only at $f_r/4$ rate, effectively dividing the mode spacing and repetition rate of the signal and idler combs.

*2.1 Repetition rate division concept with multi-seeding*

The multi-seeding approach can be generally understood both in the frequency and time domains. Let us start with the simplest case, dual seeding: When a second phase coherent CW seed laser is introduced in the system, *new signal and idler comb teeth are generated around each seed frequency* (Fig. 1b). If one sets the frequency separation between the two seed lasers to exactly half of the pump laser mode spacing $f_r/2$, the resulting mode spacing of the signal and idler combs is divided by a factor of 2. In the Results section, we demonstrate that in this case every second signal and idler pulse is removed from the pulse train as schematically shown in Fig. 1b. When the frequency separation between the two seed lasers is set to an *integer fraction of $f_r$ other than half*, the resulting frequency spacing is *still divided by that integer* *fraction*. This happens due to the additional frequency mixing during the dual-seeded OPG process. In time domain, the intensities of signal and idler pulses undergo sinusoidal modulation because of the different phase relations between the pump and seed at different times (Fig. 1c).

Finally, one can seed the OPG with multiple phase coherent lasers. In practice, this can be conveniently realized using a CW laser combined with a fast intensity modulator. In the frequency domain, the modulator produces multiple sidebands that act as multiple phase-coherent seeding sources. In the time domain, this configuration is essentially a pulse picker implemented within the OPG process: If the OPG pump power is kept below the threshold of unseeded OPG, the signal and idler pulses are generated only when the seed and pump pulses are present in the crystal at the same time (see Fig. 1d). Note that this configuration allows for repetition rate division by an arbitrarily high integer factor, which is demonstrated in the Results section below.

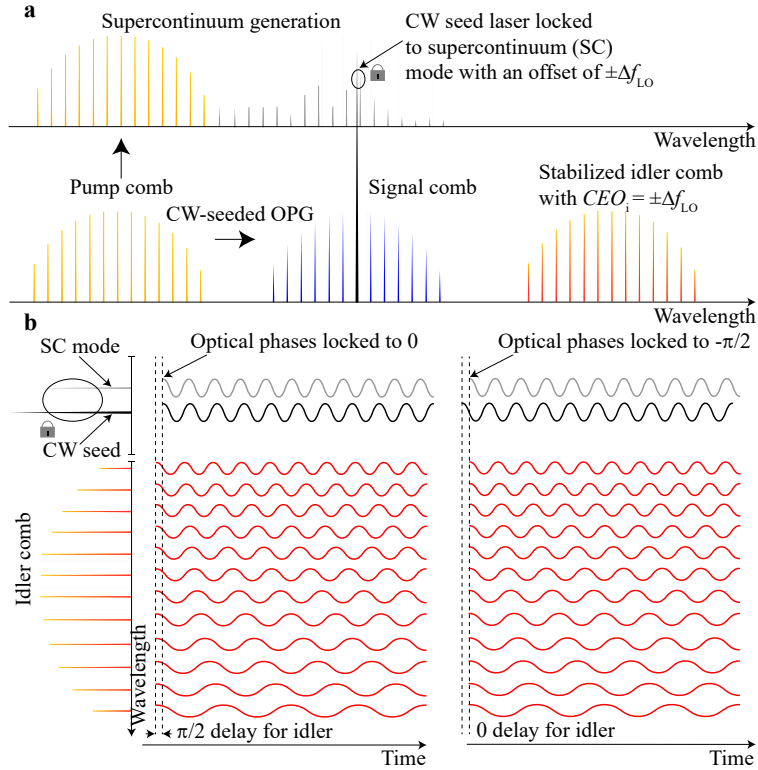

Fig. 2. (a) CEO stabilization scheme for CW-seeded or multi-seeded OPG using SC generation and phase locking. (b) Phase relations between the SC (or pump laser), signal and idler in seeded OPG.

*2.2 CEO control*

Stabilization and tuning of the CEO was demonstrated in our previous publications [32, 33], but here we briefly discuss it for completeness. To stabilize the idler comb CEO, the CW seed laser has to be referenced to the pump. This can be done if one generates suitable supercontinuum (SC) from the pump laser that reaches the CW seed laser wavelength as schematically shown in Fig. 2a. The CW seed laser is then phase locked to one of the comb teeth of the SC. The relative stability between the seed and pump leads to the *absolute* stability of the idler CEO. The frequency offset between the SC and seed is set by a radio frequency (RF) local oscillator (LO), which allows for precise adjustment of the signal and idler CEOs. This can be expressed by the following equation [32, 33]:

$$f_{CEO,p} = f_{CEO,s} + f_{CEO,i}$$
$$= (f_{CEO,p} \pm \Delta f_{LO}) + f_{CEO,i} \quad (3)$$
$$f_{CEO,i} = \mp \Delta f_{LO}$$

Here $f_{CEO,p}$, $f_{CEO,s}$ and $f_{CEO,i}$ are the carrier-envelope offset frequencies of the pump, signal and idler combs and $\Delta f_{LO}$ is the aforementioned frequency offset between the SC and CW seed. Below we confirm that this CEO stabilization technique is compatible with the repetition rate division using multi-seeding.

*2.3 Common phase control*

Referring to equation 1, the common phases of the pump, signal and idler combs are related to each other as follows [44, 45]:

$$\varphi_p = \varphi_s + \varphi_i + \frac{\pi}{2} = (\varphi_p \pm \Delta\varphi_{LO}) + \varphi_i + \frac{\pi}{2}$$
$$\varphi_i = \mp \Delta\varphi_{LO} - \frac{\pi}{2} \quad (4)$$

Here, the latter expression applies to our phase locking scheme, where the optical phases of the pump and CW seed are kept constant relative to one another (see Fig. 2b). Since the signal comb follows the seed, this leads to $\varphi_s = \varphi_p + \Delta\varphi_{LO}$, where $\Delta\varphi_{LO}$ is the phase offset of the RF local oscillator used for phase locking. In addition to the frequency offset $\Delta f_{LO}$, also $\Delta\varphi_{LO}$ can be varied, thus enabling precise adjustment of the signal comb common phase. According to Eq. (4), such adjustment is transferred to idler comb phase shift with an equal magnitude but opposite sign. This is illustrated in Fig. 2b, which schematically shows the seed laser and idler comb phases relative to the pump for two example cases. When $\Delta\varphi_{LO} = 0$, the idler phase is delayed by $\pi/2$ relative to the pump. However, when $\Delta\varphi_{LO} = -\pi/2$, the seed is delayed by quarter of its wavelength, which according to Eq. (4) leads to an idler phase delay of 0. In time domain, this means that common phase adjustments can slightly delay the signal and idler pulses relative to the pump pulse train.

### 3. Results

*3.1 Dual-seeding*

We carried out an experimental demonstration of the dual-seeded OPG using the setup that is outlined in Fig. 3 and described in more detail in the Appendix A. The pump source was a Yb-doped fiber MLL (MenloSystems GmbH, Orange comb FC1000-250) that emits 100 fs pulses at 250 MHz repetition rate. The pump beam was spatially overlapped with the CW seed laser beam(s) and focused into a MgO:PPLN crystal. The generated signal and idler beams were separated from the residual pump beam using dichroic optics and collected for analysis. The central wavelengths of the pump, signal and idler spectra were approximately 1.04 µm, 1.55 µm, and 3.34 µm, respectively.

The pump CEO was not stabilized, but the repetition rate was locked to a GPS-disciplined RF oscillator (MenloSystems GmbH, GPS-8). Two External Cavity Diode Lasers (ECDLs; Toptica Photonics, CTL 1550) were used as CW seed lasers. Their wavelengths are tunable between 1510 and 1630 nm, thus overlapping with the OPG signal spectrum. For the dual-seeding experiment, ECDL2 was phase-locked to ECDL1 such that their relative frequency separation was adjusted to a fraction of the pump repetition rate ($f_r/2$, $f_r/3$, $f_r/4$ …). The RF LO used in phase locking was referenced to the same GPS-disciplined oscillator as the pump MLL, allowing us to synchronize the optical beating between the two seed lasers with the pump repetition rate.

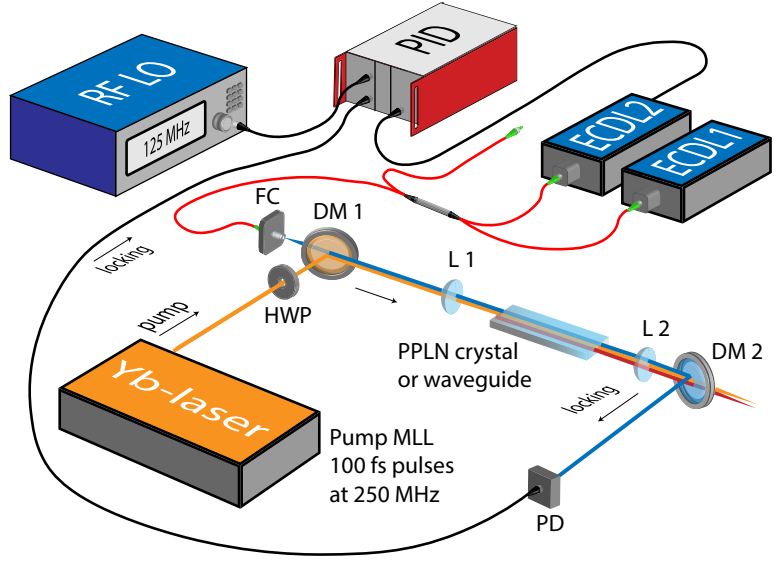

Fig. 3. Simplified dual-seeded OPG setup. ECDL2 is phase locked to ECDL1 with frequency separation of 125 MHz (half of the pump repetition rate). The pump repetition rate and RF LO are referenced to GPS-disciplined RF oscillator. Note that only essential components are shown in the figure for simplicity. FC: fiber collimator, HWP: half-wave plate, DM: dichroic mirror, L: lens, PD: photodiode, PID: proportional–integral–derivative controller, RF LO: radio frequency local oscillator, ECDL: external cavity diode laser.

We measured the generated signal comb using a high-resolution Fourier transform infrared spectrometer (FTIR; Bruker, IFS 120 HR). An interferogram for the single-seeding case is shown in Fig. 4a for comparison. Here, we resolve 3 interferometric bursts corresponding to the linear auto-correlation (interference of a pulse with itself), the first cross-correlation (pulse interferes with the first subsequent pulse) and the second cross-correlation (pulse interferes with the second subsequent pulse). Note that the drop in intensity of the second cross-correlation is an artifact caused by imperfect alignment of our FTIR at long scanning distances. For reference, we also measured the RF spectrum of the signal beam with a fast photodiode that was coupled to an RF spectrum analyzer. The RF spectrum shows a strong beat note at 250 MHz (Fig. 4b), which corresponds to the pump laser repetition rate.

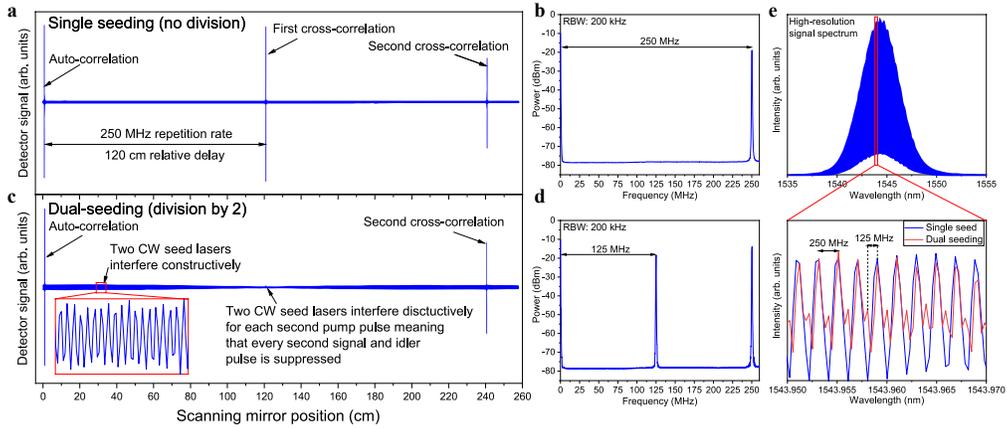

Fig. 4. (a) Interferogram of the signal pulse train with single CW seed; (b) RF spectrum of the signal pulse train shown in a; (c) Interferogram of the signal pulse train in the dual-seeded OPG when seeds are phase locked with the frequency separation of $f_r/2$ =125 MHz. Note that both the seeding and signal fields are present in the measurement; (d) RF spectrum of the signal field free of the seeding field. Since it is free of the seeding field, the 125 MHz beat note indicates the new mode spacing of the signal field; (e) High-resolution spectrum of the signal comb (free of the seeding field) with single seed (blue) and dual-seeding (red) confirming the mode spacing division by a factor of two.

To demonstrate the effect of dual seeding, we first locked the CW seed laser frequency separation to $f_r/2 = 125$ MHz. When both CW seeds are introduced in the nonlinear crystal together with the pump pulses, the average power of the signal and idler drops, reaching its minimum when the optical beating between the two seed lasers is precisely synchronized with the timing of the pump pulses. This synchronism is achieved by adjusting the RF LO phase that controls the relative optical phases between ECDL1 and ECDL2. If the pump power is lower than the OPG threshold without seeding, the average power of the signal and idler drops by a factor of two, because every second signal and idler pulse is removed from the pulse train. Next, we slightly spatially misaligned the seeding beam relative to the generated signal beam, which allowed us to measure the RF beat note spectrum and the high-resolution optical spectrum of the *signal field alone*. From the RF spectrum, shown in Fig. 4c, it is clear that the mode spacing of the signal comb is divided by a factor of 2. This mode-spacing division was also verified by measuring the FTIR spectrum of the signal comb (Fig. 4e). The same division factor was observed for the MIR idler comb.

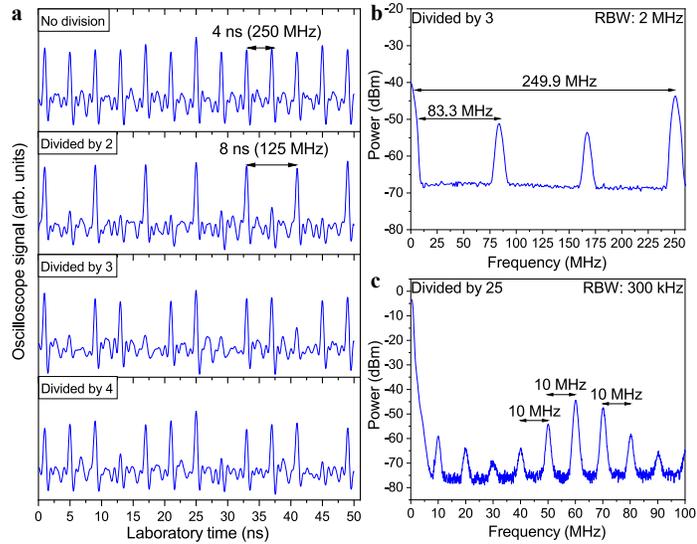

Fig. 5. (a) Oscilloscope traces in different configurations of dual-seeded OPG; (b) RF spectrum of the signal comb in the division by 3 case; (c) RF spectrum of the signal comb in the division by 25 case.

Last but not least, we measured the signal pulse train using a fast oscilloscope. These time-domain results are illustrated in Fig. 5a for three different division factors, corresponding to CW seed laser frequency separations $f_r/2$, $f_r/3$, and $f_r/4$. The single-seeded case is also shown for comparison. For a division factor M, each $M^{th}$ pulse is removed from the pulse train. In the frequency domain, this corresponds to comb mode-spacing division by M. This is exemplified by the RF spectra of M = 3 and M = 25, shown in Figs. 5b and b, respectively. The efficiency of the additional optical frequency mixing is limited, so for high division factors the newly generated comb lines are typically less intense than the lines corresponding to the original mode spacing.

*3.2 Multi-seeding*

The experimental setup used to demonstrate the multi-seeding approach is outlined in Fig. 6. For most parts, the setup is the same as that used in dual-seeding experiments. Instead of two separate CW lasers, however, we used a single intensity-modulated CW laser to produce seeding at multiple optical frequencies. The figure also shows the SC generation part that was needed for CEO stabilization, which is discussed in the following section. We modulated the CW seed laser (ECDL in Fig. 6) with an electro-optic intensity modulator (EOM; Lucent Technologies, 2623Y) that was driven with an RF pulse generator (PG; Keysight, 81150A).

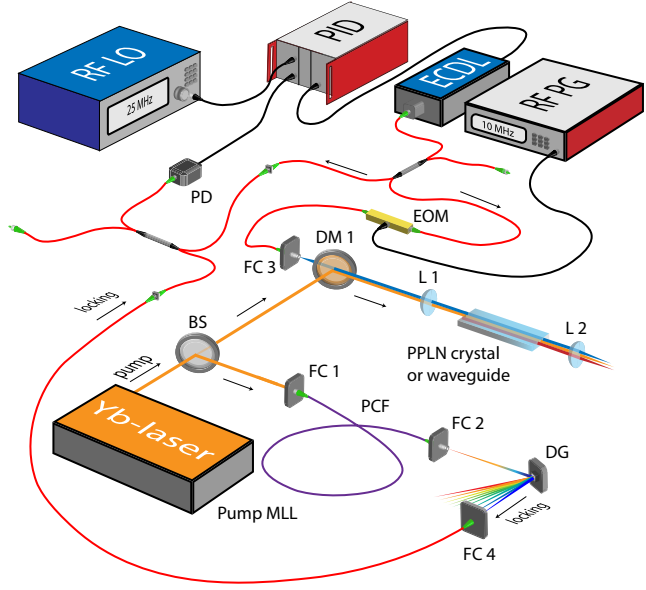

Fig. 6. Multi-seeded OPG setup. ECDL is split in two arms, one is used for phase locking to SC and the other for intensity modulation in EOM. The intensity modulated output is then used for seeding. The pump repetition rate, RF PG and RF LO are referenced to GPS-disciplined RF oscillator. Note that only essential components are shown in the figure for simplicity. FC: fiber collimator, HWP: half-wave plate, DM: dichroic mirror, L: lens, PD: photodiode, PID: proportional–integral–derivative controller, RF LO: radio frequency local oscillator, RF PG: radio frequency pulse generator, ECDL: external cavity diode laser, PCF: photonic crystal fiber.

The PG was referenced to the same GPS-disciplined RF oscillator as the pump laser, allowing us to precisely synchronize the seed laser modulation with the pump laser repetition rate.

The electrical pulses generated by the RF PG are shown in Fig. 7a (top) for the shortest available pulse duration of 8.5 ns. When these pulses are used to drive the EOM, we get seed laser pulses with a freely tunable pulse repetition rate (Fig. 7a (bottom)). For demonstration purposes, we set the seed modulation rate to 10 MHz, resulting in a comb repetition rate division factor of $M = f_r/(10\ \mathrm{MHz}) = 25$. The signal pulse train was measured with a fast photodiode and an oscilloscope, as is shown in Fig. 7b. Comparison with the unmodulated signal pulse train (Fig. 7b top) reveals the expected repetition-rate division by 25. The respective mode-spacing division in frequency domain is apparent in the RF spectrum, Fig. 7c. In this case, all the harmonics in the RF spectrum have almost the same power, unlike in the dual-seeded case shown in Fig. 5c. The same repetition-rate and mode-space divisions are observed also for the MIR idler comb, as demonstrated in Figures 7d and e.

With the aforementioned settings, we also measured Relative Intensity Noise (RIN) of the MIR comb (Fig. 8). The measured integrated RIN is as low as 0.058 %, which is essentially the same value as that of single-seeded non-saturated OPG, 0.053 %, measured for similar conditions [32].

*3.3 CEO control*

To prove that our previously demonstrated offset frequency control technique [32, 33] is compatible with the multi-seeding approach introduced here, we characterized the idler comb CEO the same way as we did previously [32]. This was performed using the setup of Fig. 6, but with the bulk MgO:PPLN crystal replaced by a MgO:PLLN waveguide [33]; see Appendix A for details. First, we phase-locked the CW seed laser to the pump laser SC generated in a photonic crystal fiber (Fig. 6). This scheme leads to idler CEO stabilization as explained in the Principle section. Then, the seed laser beam was passed through the intensity modulator, which was configured for repetition rate division by $M = 10$; see the previous paragraph. In other words, we produced an idler comb with $f_r/M = 25$ MHz mode spacing.

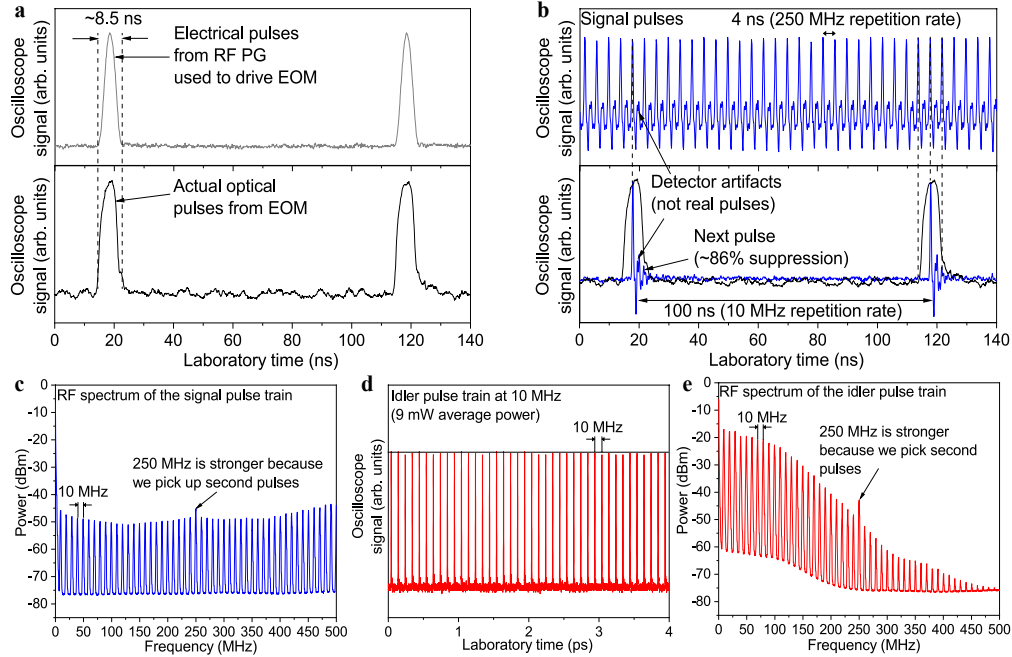

Fig. 7. (a) Electrical pulses from RF PG (top), and the corresponding optical pulses generated in EOM (bottom). (b) Signal pulse train with single seeding (top) and multi-seeding (bottom). Note that the bottom trace is free of the seeding field, but we superimposed the seeding pulses (black curve) to guide the eye. (c) RF spectrum that corresponds to the oscilloscope trace shown in b (bottom). (d) Oscilloscope trace of the idler pulse train in multi-seeded OPG. (e) The corresponding RF spectrum of the idler pulse train shown in d. The detector bandwidth is about 120 MHz.

To confirm the stability and tuning of the MIR idler comb CEO, we frequency-doubled the idler comb in order to compare it against another fully stabilized OFC: An Er-fiber laser comb (MenloSystems GmbH, FC1500-250-WG) that was referenced to the same GSP-disciplined RF oscillator as all other instruments used in the experiment. For a comb-to-comb comparison, an auxiliary ECDL was phase-locked to a tooth of the Er-fiber laser comb, and subsequently heterodyned with the frequency-doubled idler comb to obtain coherent beat notes. These beat notes can be seen in the RF spectrum shown in Fig. 9b. The strongest beat notes represent the idler comb repetition rate of 25 MHz and the weak ones indicate the auxiliary ECDL frequency beating against the frequency-doubled idler comb. By changing the RF LO frequency of the CEO stabilization loop, we varied the signal and idler CEOs according to Eq. (3). The actual idler CEO frequency was then tracked by observing the beat notes of the idler second harmonic against the auxiliary ECDL, as exemplified in Fig. 9c. Since the meaningful CEO tuning range is defined by the idler comb mode spacing, the CEO counting starts from zero after every $f_r/M$ = 25 MHz.

*3.4 Common phase control*

In order to demonstrate precise tuning of the common phases $\varphi$ (Eq. (1)) of the OPG output combs, we set up an additional measurement scheme for that of the signal comb. A detailed description of this scheme can be found in Appendix B. In brief, we used the multi-seeding setup of Fig. 6 with the CW seed laser (ECDL) phase-locked to the pump SC at 1549.6 nm wavelength. An auxiliary ECDL was then phase-locked to the same SC but at a different wavelength, allowing us to track the phase of a signal comb tooth far away from the seeded modes. The tracking was carried out at three different wavelengths (1547, 1550 and 1552 nm) by heterodyning the signal comb with the auxiliary laser on a fast photodiode. The heterodyne signal was fed into an RF mixer that served as a phase detector – that is, we monitored the relative phase difference between one of the signal comb teeth and the auxiliary laser by measuring the DC output voltage of the mixer. Next, we gradually tuned the common phase of the signal comb by adjusting $\Delta\varphi_{\text{LO}}$, *i.e.* the phase of the RF local oscillator that was used to lock the seed laser (see Eq. (4)). Results of the measurements can be seen in Fig. 9d for the three reference wavelengths. It is evident that the optical phase can be continuously adjusted and the adjustment is identical regardless of the comb tooth under investigation.

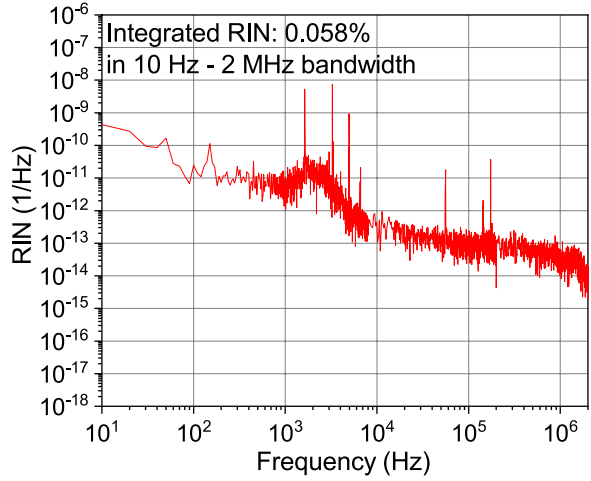

Fig. 8. Idler RIN in the multi-seeded OPG case.

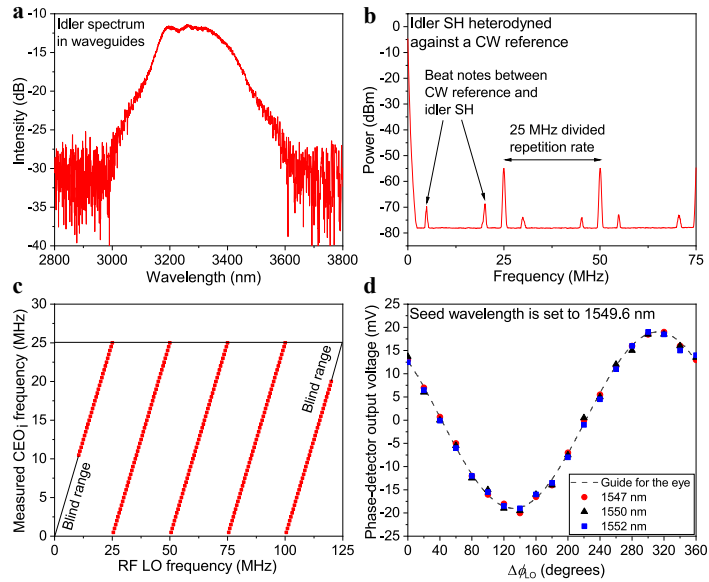

Fig. 9. (a) Typical idler spectrum in our waveguide system. (b) RF spectrum of the idler second harmonic comb heterodyned against a CW reference. (c) Idler CEO tuning demonstration for the divided repetition rate of 25 MHz. (d) The common phase measurement, see text for details.

## 4. Conclusion

To summarize, we have demonstrated a method for versatile optical frequency synthesis based on multi-seeded OPG that allows adjusting three major degrees of freedom – repetition rate, CEO and common phase. Our method is general and it is not limited to the implementation we describe in this article. Hence, a variety of pump and seed lasers as well as intensity modulators and nonlinear materials can be used.

Although the dual- and multi-seeding methods are similar, they can be used in different situations. The multi-seeding approach is useful when large repetition rate division factors (>10) are required. In the pulse picking configuration it can only be used if the RF pulse generator and the EOM are fast enough to pick single pulses. Hence, in practice, it is suitable for initial repetition rates <1GHz or pulse intervals >1 ns. On the other hand, the dual-seeding is applicable to relatively small division factors (<10) because efficiency of the additional frequency mixing is limited. For dual-seeding, however, there are no additional limitations for the initial repetition rate, so in principle it can be used even if the time interval between subsequent pulses is very small.

As for the applications, we anticipate that this technique can be useful in dual-comb based techniques for gas phase molecular spectroscopy [31, 46-48]. In addition, OPG can be used as a source of squeezed light [49] and entangled photon pairs [50, 51]. For instance, CW-seeded OPG has recently been used as a source of unheralded photons for single photon interferometry [52]. However, the current state of single photon detectors does not allow detection at high frequencies [40], hence the repetition rate division is a useful addition. Another application example is optical chirped-pulse amplification, where generation of high pulse energy often requires substantial downscaling of the pulse repetition rate [53-55]. On the other hand, future advances in nanophotonic waveguide technologies [56-58] may allow us to drive OPG with ultra-low energy pulses (sub-femtojoules), which would offer possibilities for high repetition rate frequency combs to be used as pump sources, potentially including soliton microcombs. Hence, a flexible repetition rate division would greatly complement such systems.

**Appendix A: Experimental setup**.

We used MgO-doped periodically poled lithium niobate (MgO:PPLN) as the nonlinear medium. Most of the experiments were done using a 50 mm long bulk MgO:PPLN fanout crystal with poling periods of 26.5 µm–32.5 µm (HC Photonics) that was placed in an oven with a constant temperature of 100 ºC. This implementation takes advantage of so-called pulse trapping effect [59] and allows us to generate tunable signal (1.5 W maximum average power) and idler (700 mW average power) OFCs in the 1400 nm – 1560 nm and 3325 nm – 4000 nm spectral ranges, respectively [32]. Besides MgO:PPLN [60-62], OPG with and without seeding has been demonstrated in many different nonlinear materials [63-67], which helps to generate wavelengths unattainable in MgO:PPLN.

In the CEO and common phase tuning demonstrations, we used a waveguide made of MgO:PPLN (HC Photonics). The waveguide is 12 mm long, 5.6 μm wide and 4.3 μm high with a poling period of 21.71 μm. The waveguide was placed in an oven with a constant temperature of 75 ºC. This system has an extremely low threshold of just 25 pJ for 100 fs pump pulses at 250 MHz repetition rate [33]. The maximum average powers of the signal and idler combs are 17.4 mW and 5.8 mW, respectively. Idler second harmonic (SH) is generated directly in the waveguide with a peak at ~1611 nm, so there was no need for an external frequency doubling scheme: The idler SH conveniently overlaps with the wavelength range of our ECDLs (1510 - 1630 nm), enabling the CEO tuning experiment reported in Fig. 9c.

For optimal operation of the multi-seeded OPG, it is important to set the pump power below the threshold of the OPG without seeding. This minimizes the generation of unwanted signal and idler pulses or, in other words, it helps to start the OPG process only when the seed and pump pulses overlap in time. The seeding field decreases the threshold by up to 30 % in the bulk crystal implementation (50% for the waveguide). For the case presented in Fig. 7, we set the average pump power to 3.2 W, which is significantly lower than the average power required to start OPG without seeding (4.3 W for ~1545 nm signal and ~3350 nm idler phase matching condition). The seeding average power was set to 10 mW (at 10 MHz repetition rate and ~8.5 ns pulse duration). With the abovementioned settings, the average output idler power measured after a germanium window was about 8 mW. This power level corresponds to 200 mW of average idler power without division. However, if one increases the pump power above 3.2 W (despite it being below the threshold of the OPG without seeding), due to the increased optical parametric fluorescence at higher pump powers, some unwanted pulses start to be noticeable. Essentially, the optical parametric fluorescence is the spontaneous radiation generated below the OPG threshold without much amplification. Hence, to minimize this effect, the pump power should be significantly lower (25% or more) than the OPG threshold for OPG without seeding. Note also that because the minimum pulse duration available from our RF PG was ~8.5 ns, the second consecutive pulse cannot be completely removed from the pulse train. However, nowadays one can have RF PGs with pulse duration as short as 4 ns [39] meaning that it can be possible to remove these unwanted pulses from the pulse train even at the initial 250 MHz repetition rate and beyond.

Regarding the CEO tuning and stability, the details on CEO stabilization and phase-locking procedure have been described previously [32, 33]. The blind ranges of the CEO tuning curve shown in Fig. 9c correspond to idler CEO frequencies that we were unable to attain due to the limited range of input frequencies of the locking electronics (Toptica Photonics, mFALC 110) that we used to phase-lock the CW seed laser to pump supercontinuum.

**Appendix B: Common phase measurements**.

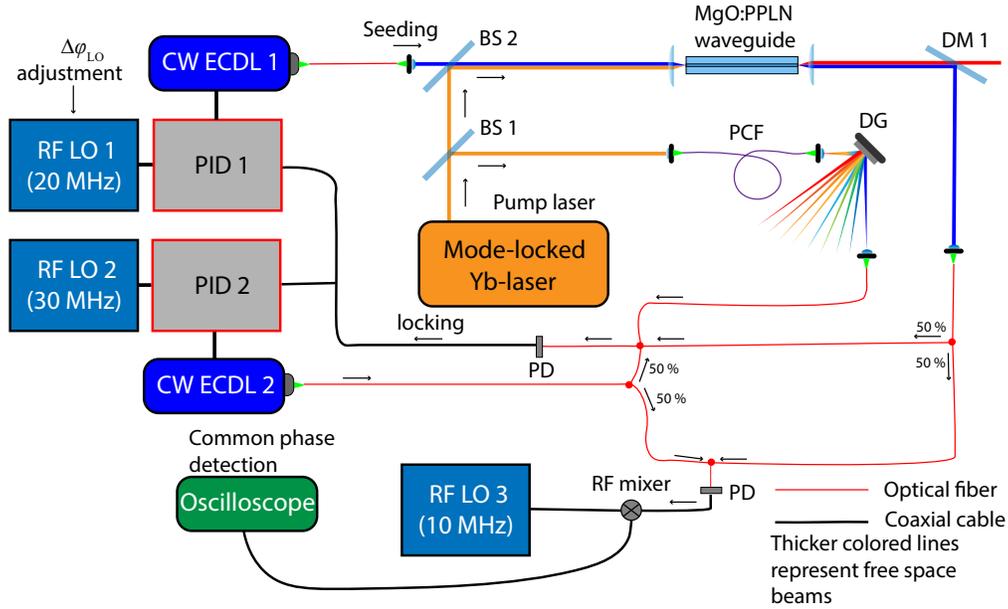

Fig. 10. Scheme for the common phase measurements. DM: dichroic mirror, PD: photodiode, PID: proportional–integral–derivative controller, RF LO: radio frequency local oscillator, ECDL: external cavity diode laser, PCF: photonic crystal fiber, DG: diffraction grating.

In Fig. 10 one can see the experimental setup for the common phase measurement. For simplicity, this setup does not include the repetition rate division part. The first CW external cavity diode laser (ECDL1 at 1549.6 nm) was used to seed the OPG in a MgO-doped periodically poled lithium niobate (MgO:PPLN) waveguide. The coupling of the pump and seeding beams into the waveguide was done in free space. At the output of the waveguide we picked up the generated signal comb and coupled it into an optical fiber. A portion of the pump power (~160 mW of average power) was used to generate a supercontinuum. A diffraction grating was used to filter part of the supercontinuum near 1550 nm (~10 nm bandwidth). The filtered part of the supercontinuum was coupled into another optical fiber, combined with the light from ECDL1, and detected with a fast photodetector. The photodetector signal was processed in a phase detector that comprised a radiofrequency local oscillator (RF LO), mixer and filters [32]. The phase detector output was used as an error signal for a proportional-integral-derivative controller to phase-lock ECDL1 (seed) to the pump supercontinuum for CEO stabilization. The CEO and common phase of the signal comb could thus be controlled by adjusting the frequency and phase of the local oscillator, see equations (3) and (4) of the main text.

The second CW external cavity diode laser (ECDL2) was used as an optical phase reference that allowed us to track the phases of signal comb modes upon tuning the phase of the seed laser, ECDL1. In order to do that, we phase-locked ECDL2 to the supercontinuum the same way as we did for ECDL1. We used separate RF local oscillators, RF LO1 and RF LO2, respectively, for the phase locks of ECDL1 and ECDL2. Both local oscillators were however referenced to the same GPS-disciplined oscillator. Once the phase locks were established, we combined part of the signal comb with ECDL2 light in another optical fiber and sent them to a second photodiode. Using this photodiode, we detected the beat note between the signal comb and the reference laser. This beat note was located at 10 MHz, corresponding to the difference frequency between the two RF LOs (30 MHz – 20 MHz). Finally, the beat note phase was detected with a mixer, which had as a reference a third RF LO (also reference to the same GPS-disciplined oscillator). The phase was tracked by monitoring the DC level of the mixer output with an oscilloscope.

During the measurements, the phase of RF LO 1 was varied in order to vary the optical phase of ECDL1 relative to the pump comb. (Recall that the ECDL1 phase defines the common phase of all the signal/idler comb lines according to Eq. (4)). On the other hand, the phase of ECDL2 (reference laser) was kept constant by maintaining a constant phase of RF LO 2. As a result, the DC output of the mixer changes as demonstrated in Fig. 9d. This was confirmed for three different wavelengths of the reference laser.

**Disclosures.** The authors declare no conflicts of interest.

**Data availability.** Data underlying the results presented in this paper are not publicly available at this time but may be obtained from the authors upon reasonable request.

**Acknowledgment.** The authors wish to thank Juho Karhu and Santeri Larnimaa from the University of Helsinki, Mikko Närhi from Tampere University, as well as Thomas Fordell from VTT Technical Research Centre for valuable discussions on the manuscript. We also thank Jui-Yu Lai from HC Photonics (Taiwan) for providing the nonlinear PPLN waveguides. This research was supported by the Flagship on Photonics Research and Innovation of the Academy of Finland (PREIN). Mikhail Roiz acknowledges financial support from the Alfred Kordelin foundation.